\documentclass[twocolumn,superscriptaddress,aps,showpacs,floatfix,prl,10pt]{revtex4-1}
\usepackage[T1]{fontenc}
\usepackage{amssymb,amsmath,bm}
\usepackage{graphicx}
\usepackage{color}
\usepackage{bbold}
\usepackage{bbm}
\usepackage{appendix}
\usepackage{soul}
\usepackage[dvipsnames]{xcolor}
\usepackage{ulem}
\usepackage{float}
\usepackage{textcomp}
\usepackage{ mathrsfs }
\usepackage{siunitx}%

\begin{document}
\title{The Roles of Kerr Nonlinearity in a Bosonic Quantum Neural Network}
\author{Huawen Xu}
\email{huawen001@e.ntu.edu.sg}
\affiliation{Division of Physics and Applied Physics, School of Physical and Mathematical Sciences, Nanyang Technological University, 21 Nanyang Link, Singapore 637371, Singapore}
\author{Tanjung Krisnanda}
\affiliation{Division of Physics and Applied Physics, School of Physical and Mathematical Sciences, Nanyang Technological University, 21 Nanyang Link, Singapore 637371, Singapore}
\author{Ruiqi Bao}
\affiliation{Division of Physics and Applied Physics, School of Physical and Mathematical Sciences, Nanyang Technological University, 21 Nanyang Link, Singapore 637371, Singapore}
\author{Timothy C. H. Liew}
\email{timothyliew@ntu.edu.sg}
\affiliation{Division of Physics and Applied Physics, School of Physical and Mathematical Sciences, Nanyang Technological University, 21 Nanyang Link, Singapore 637371, Singapore}
\affiliation{MajuLab, International Joint Research Unit UMI 3654, CNRS, Universit\'e C\^ote d'Azur, Sorbonne Universit\'e, National University of Singapore, Nanyang Technological University, Singapore}

\begin{abstract}
The emerging technology of quantum neural networks (QNNs) attracts great attention from both the fields of machine learning and quantum physics with the capability to gain quantum advantage from an artificial neural network (ANN) system. Comparing to the classical counterparts, QNNs have been proven to be able to speed up the information processing, enhance the prediction or classification efficiency as well as offer versatile and experimentally friendly platforms. It is well established that Kerr nonlinearity is an indispensable element in a classical ANN, while, in a QNN, the roles of Kerr nonlinearity are not yet fully understood. In this work, we consider a bosonic QNN and investigate both classical (simulating an XOR gate) and quantum (generating Schr\"{o}dinger cat states) tasks to demonstrate that the Kerr nonlinearity not only enables non-trivial tasks but also makes the system more robust to errors.
\end{abstract}
\maketitle
\emph{Introduction.}--Biologically inspired artificial neural networks (ANN) have shown great accomplishments in processing information with the ability to break through the von Neumann bottleneck (referring to the delay between processor and memory)~\cite{Jelena2015,Chouard2015,LeCun2015,Butler2018}. Various types of ANN architectures have been proposed in the field of machine learning, for example, feedforward neural networks~\cite{Bebis1994} and recurrent neural networks~\cite{LeCun2015,Herbert2004,Maass2002,Enel2016,Lukosevicius2009} that are shown to be powerful in speech/pattern/fingerprint classification~\cite{Deng2013,Chan2016,Ciresan2011,Leung1991}, financial forecasting~\cite{Kaastra1996} and nonlinear series prediction~\cite{Zhang2001}. We normally aim for more complexity from an ANN to obtain richer dynamics. This in turn allows for a better performance for the tasks mentioned above. For example, this can be achieved by considering more layers/nodes~\cite{Xu2017}, more completed architectures~\cite{Liu2016,Zhang2019} and stronger nonlinearity~\cite{Maas2013}, all contribute to performance.

Based on these requirements of an ANN, many physical systems have been proposed to be hardware implementation platforms for ANNs, for example: memristors~\cite{Krzysteczko2012}, spintronics ~\cite{Locatelli2014,Quang-Diep2014}, microcavity exciton-polaritons~\cite{Xu2020,Opala2019,Ballarini2020,Matuszewski2021}, etc. Among these platforms, a nonlinear activation function is an indispensable part for even basic machine learning tasks. The role of the activation function is to determine the output of a specific node given a set of corresponding inputs. This means that if we allow only a linear activation function in an ANN, the outputs will simply be a linear transformation of the inputs (no matter how many network layers one implements). Such an ANN can be represented just by a matrix multiplication corresponding to a simple input-to-output process. Nonlinearity is required from a physical system for hardware implementation for more general transformations. 
 
Recently, quantum neural networks (QNNs)~\cite{Biamonte2017,Dunjko2018,Altaisky2016} emerged as promising platforms combining the characteristics of ANN and quantum physics, which aim to gain advantages of quantum mechanics, for example, having more degrees of freedom from the large Hilbert space~\cite{McClean2018} and quantum correlations between quantum modes~\cite{Shen2020} in performing either classical or quantum tasks. In this direction, QNNs have been shown to offer speedup in solving classical tasks~\cite{Szegedy2004, F2014,Dunjko2016,Paparo2014} and improvement in learning efficiency~\cite{Xu2021,Nakajima2019,Fuji2017,Neigovzen2009}, compared to their classical counterparts. On the other hand, quantum tasks such as entanglement recognition~\cite{Ghosh2019a}, phase estimation~\cite{krisnanda2021beating}, quantum state preparation~\cite{Krisnanda2021,Ghosh2019} and quantum state tomography~\cite{Ghosh2021} have also been proposed with different kinds of QNNs. 

Intuitively, one expects that nonlinearity is also important in a QNN similar to the classical ANN, which requires further demonstration. In this work, we take a bosonic QNN as an example with different architectures and simple nearest neighbour hopping between the quantum network nodes. The architectures shall be kept simple as the aim is to demonstrate how the Kerr nonlinearity plays a role in the QNN for both classical and quantum tasks. We first choose the classical task as simulating an XOR gate, where having nonlinearity in the input-to-output process is necessary. One should note that here both the input signal and output signal are classical. We show that Kerr nonlinearity indeed allows for the XOR gate to function. It is worth noting that this is done without nonlinear elements from anywhere else. The latter can emerge simply from considering different quantities for the inputs and outputs, for example amplitude and intensity, which in turn clouds the roles of Kerr nonlinearity. Moreover, by considering a more realistic situation where the measurement error is unavoidable in practice, we found that the Kerr nonlinearity has an error correction effect for classical tasks. In particular, we consider errors on the measured outputs, and found that stronger Kerr nonlinearity leads to reduction of errors for the XOR gate. Next, we extend this investigation to the quantum regime where we consider a quantum operation generating a corresponding Schr\"{o}dinger cat state given an arbitrary coherent state. {Here the outputs are quantum states. Since a Schr\"{o}dinger cat is usually characterized by its Wigner function, we choose the cost function to minimize the difference between the Wigner function of the obtained states and target Schr\"{o}dinger cat states. We show that Kerr nonlinearity plays an essential role that allows this quantum operation and makes the QNN noise/error resistant.}

\emph{The model.} The considered quantum neural network (QNN) consists of $n$ bosonic modes with random nearest neighbour coupling. The Hamiltonian can be expressed as:
\begin{align}
\mathcal{\hat H}_0 =\sum_{i=1}^{n}\left( E \hat a_i^{\dagger} \hat a_i  +\alpha \hat a_i ^\dagger \hat a_i ^\dagger \hat a_i \hat a_i \right) +\sum_{ij} J_{ij}(\hat a_i \hat a_j^{\dagger} + \hat a_i^{\dagger} \hat a_j),
\label{Equation1}
\end{align}
where we consider an onsite energy $E$ and strength of Kerr nonlinearity $\alpha$ the same for each bosonic mode. Also, $J_{ij}$ represents the nearest neighbour coupling strength between modes $i$ and $j$.
In our scheme, the QNN is evolved following the quantum master equation:
\begin{align}
i\hbar {\dot \rho}= [\mathcal{\hat H}_0, \rho] +\sum_{i=1}^n \frac{i\gamma}{2} \mathscr{L}(\rho,\hat a_i),
\label{Equation2}
\end{align}
where $\mathcal{\hat H}_0$ is the Hamiltonian in Eq.~\ref{Equation1} and the Lindblad operator $\mathscr{L}$ is defined as $\mathscr{L}(\rho,\hat x)  \equiv 2 \hat x \rho \hat x^{\dagger} - \hat x^{\dagger} \hat x \rho - \rho \hat x^{\dagger} \hat x $. The second term on the right-hand side of Eq.~\ref{Equation2} determines the decaying process of the modes in the QNN with a rate $\gamma/\hbar$. We obtain the density matrix $\rho(\tau)$ and occupation number $\left \langle n_i \right\rangle = \text{Tr} \{ \hat a_i^\dagger \hat a_i \rho(\tau)\}$ of each mode at a certain time $\tau$ from the evolution of Eq.~\ref{Equation2}.

\begin{figure}[t]
\includegraphics[width=1\columnwidth]{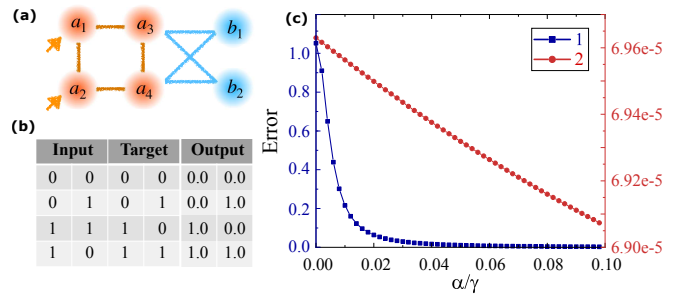}
\caption{\textbf{(a)} Schematic of a quantum neural network with nearest neighbour coupling (brown part), where $\hat a_1$ and $\hat a_2$ are the input modes and $\hat  a_{3,4}$ are the output modes. The occupation number on $\hat a_{3,4}$ are weighted to obtain the final outputs ($\hat b_{1,2}$). \textbf{(b)} Input and corresponding target outputs for a XOR gate. The results from the QNN are shown in the  "Output" columns. \textbf{(c)} Output error as a function of the strength of Kerr nonlinearity (in the unit of $\gamma$) for two input signal encoding methods. $1$ represents the occupation number input encoding scheme whereas $2$ means the coherent pumping amplitude encoding method.}
\label{Figure1}
\end{figure}

\begin{figure}[h!]
\includegraphics[width=1\columnwidth]{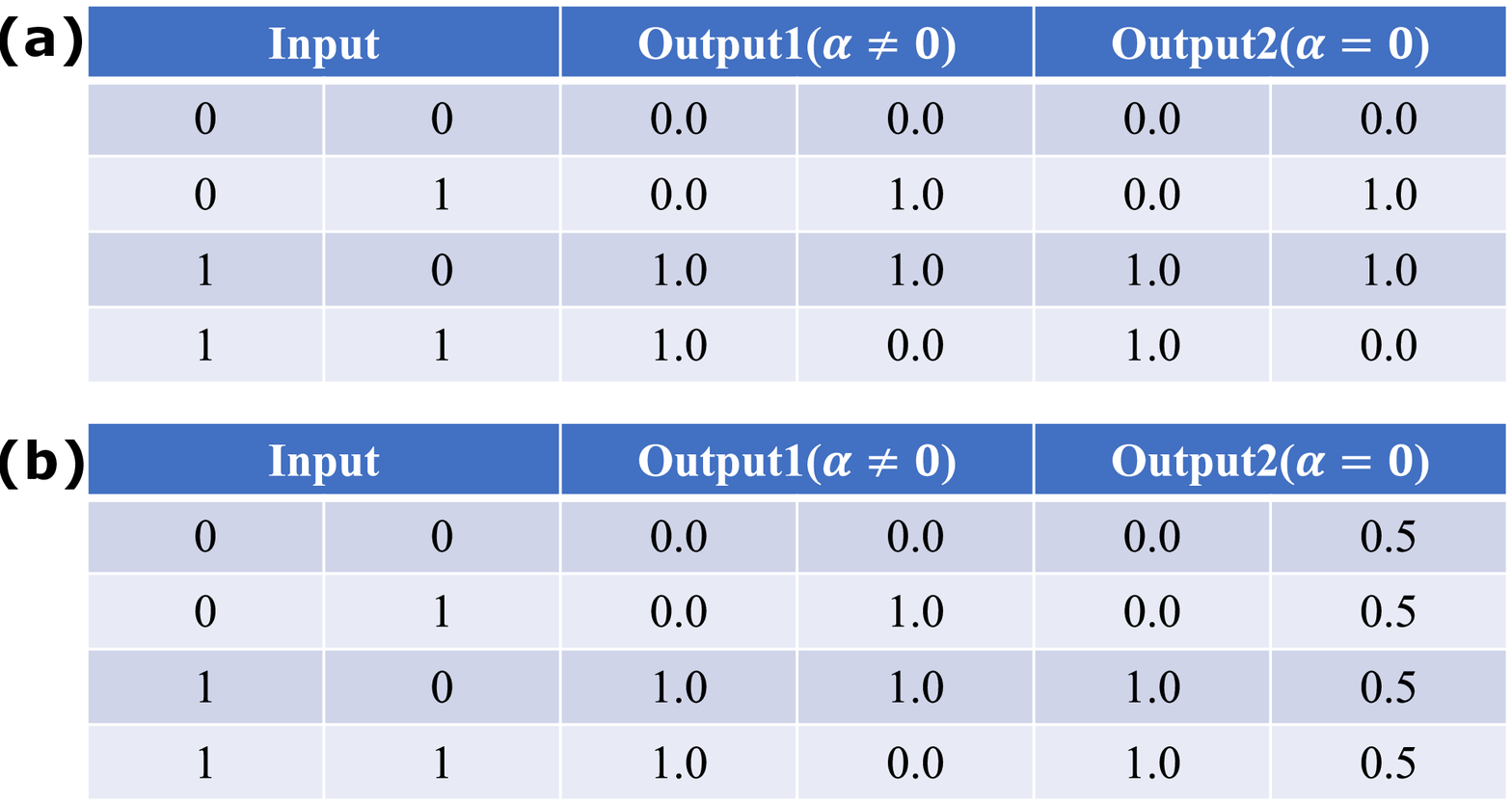}
\caption{\textbf{(a)} Input output correspondence when the intensity $|\psi_n|^2$ is measured as the outputs for cases when Kerr nonlinearity $\alpha$ is set to be non-zero (Output1) or zero (Output2). \textbf{(b)} Input output correspondence when the wavefunction amplitude $\psi_n$ is measured as the outputs for cases when Kerr nonlinearity $\alpha$ is set to be non-zero (Output1) or zero (Output2).}
\label{Figure2}
\end{figure}

\emph{Classical Task.} Here we consider the classical task of simulating a classical XOR gate with a four bosonic mode QNN; the input and target output relation is shown in Fig.~\ref{Figure1}\textbf{(b)}. The first two modes ($\hat a_{1,2}$) are the input modes and the last two modes ($\hat a_{3,4}$) are the output modes (see Fig.~\ref{Figure1}\textbf{(a)}). The input signals can be introduced as amplitudes of coherent pump or encoded as the occupation numbers for the input modes. After the evolution for a time $\tau$, we measure the occupation numbers on the output modes. We first investigate the case of injecting the input signals via coherent pump, in which the Hamiltoinian in Eq.~\ref{Equation1} now reads:
\begin{align}
\mathcal{\hat H}_0 \rightarrow  \mathcal{\hat H}_0 + \sum_{i=1}^{2}(P_i \hat a_i^\dagger + P_i^* \hat a_i),
\label{Equation3}
\end{align}
where the amplitudes of the coherent pump correspond to the binary input in Fig.~\ref{Figure1}\textbf{(b)}. This form of pumping is consistent with photonic neural network prototypes based on exciton-polaritons~\cite{Ballarini2020}. It turns out that without Kerr nonlinearity ($\alpha \hat a ^\dagger \hat a ^\dagger \hat a \hat a$), the QNN can already simulate an XOR gate, i.e., the trained outputs can match the target with negligible error, see the last two columns in the table of Fig.~\ref{Figure1}\textbf{(b)}. This somewhat unexpected result can be explained by noting that the inputs and outputs in the QNN are nonlinearly related. That is, the inputs are encoded as pumping amplitudes while the output as occupation numbers. As occupation number is a nonlinear function of amplitude, the system is capable of nonlinear input-to-output mapping corresponding to the XOR gate. Similar results are also demonstrated in Ref.~\cite{Mujal2021}. Given that the Kerr nonlinearity is not required, we also verified that an evolution of classical states treated within a mean-field model of the system can reproduce the considered XOR gate (see the results in Fig.~\ref{Figure2}), which also rules out the importance of quantum correlations in this task. Details of simulating the classical neural networks can be found in the Supplementary Materials (SM). Let us also note that if we were to instead consider particle intensities rather than pumping amplitudes as the inputs, the system would no longer be capable of simulating a nonlinear map and the XOR gate would not be possible. While the simple investigation so far suggests that Kerr nonlinearity is not necessary for the QNNs and ANNs, as one might obtain nonlinear elements in the system by other means, we will show below that it does offer an essential element in situations accounting for measurement errors, noise, or more complicated tasks.

Let us consider noise introduced on the measured occupation number of each mode. Instead of taking $\left \langle n_i \right\rangle$ as the output, $\left \langle n_i \right\rangle + \delta_i$ is considered, where $\delta_i$ is the measurement error on the $i$-th mode. We consider a uniformly distributed random error $\delta_i=[0,0.8]\times \left \langle n_i \right\rangle$. This model is such that the errors are fractions of the corresponding occupation numbers. The error of the trained outputs is defined as the difference between the targets and the actual trained outputs. Note that since in a logic gate, both the inputs and outputs are typically binary digits, the error must be below $0.5$ to operate successfully. Now, we encode the input information with occupation numbers in modes $\hat a_1$ and $\hat a_2$, which means instead of starting from vacuum state for all inputs, we start with coherent states with average occupations $0/1$ in modes $1$ and $2$ according to the different inputs in Fig.~\ref{Figure1}\textbf{(b)}. In this case, no pumping scheme is considered in the system (set $P=0$ in Eq.~\ref{Equation2}). In Fig.~\ref{Figure1}\textbf{(c)} (blue curve), we can see the error drops rapidly to insignificant as the nonlinearity increases. However, the error is non-negligible when zero Kerr nonlinear strength is considered ($>0.5$).

Now, we revisit the scheme where the inputs are encoded in the amplitudes of the coherent pump $P_{1,2}$. In this case, we start the system from vacuum states for all the modes. In Fig.~\ref{Figure1}\textbf{(c)} (red curve), we can see the error is already insignificant even when $\alpha$ is set to zero. As mentioned above, this is because there is nonlinearity already present, i.e., that between the input and output quantities. One can still see that the error drops with increasing of the strength of Kerr nonlinearity. Intensity encoded input is also consistent with neural networks based on non-resonantly pumped polariton condensates~\cite{Mirek2021}, which are considered explicitly in the SM. This concludes that Kerr nonlinearity offers an essential element for QNNs to simulate nonlinear processes such as the XOR gate, in the presence of measurement errors.
\begin{figure}[b]
\includegraphics[width=1\columnwidth]{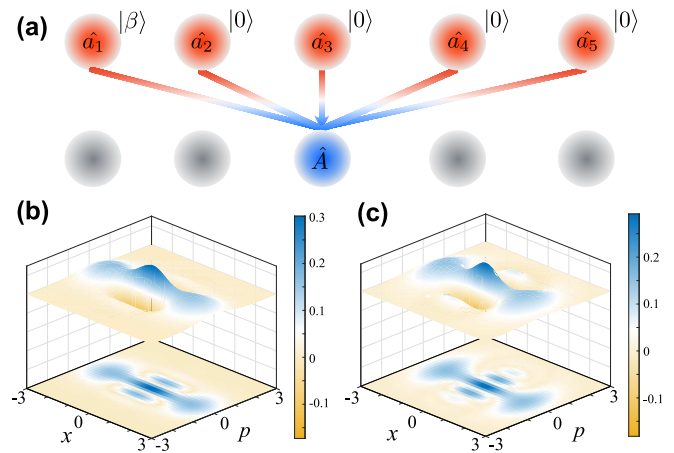}
\caption{\textbf{(a)} {Schematic of the linear mixing process of the QNN. The photons emitted from the QNN modes $\hat{a_i}$ (red parts) can be recombined to form new modes $\hat{A}$ (blue parts). Since we are targeting at one output mode, the other modes are assumed to be vacuum (gray parts).} \textbf{(b)} {The Wigner function of the target Schr\"{o}dinger cat state.} \textbf{(c)} {The Wigner function of the prepared Schr\"{o}dinger cat state with an error $\delta=0.07$. Here we consider a five bosonic mode QNN with random nearest neighbour coupling. Parameters: $P=1000\gamma, J=1000\gamma, \tau = 0.01/\gamma$, and $\alpha = 450\gamma$.}}
\label{Figure3}
\end{figure}

\emph{Schr\"{o}dinger cat state generating operation.} Here, we consider a quantum task utilising the QNN, that is, generating a Schr\"{o}dinger cat state from a given coherent state $ \left | \beta \right\rangle$. The target Schr\"{o}dinger cat state is defined as: 
\begin{align}
\left | Cat_{\beta,k}\right\rangle =\frac{1}{N_{\beta,k}}[\left | \beta \right \rangle +\left | -\beta \right \rangle],
\label{Equation4}
\end{align}
where the coherent state is expressed as $ \left | \beta \right\rangle = e^{\beta \hat a^\dagger-\beta^* \hat a}\left | 0\right \rangle$, the coefficient $N_{\beta,k} = \sqrt{2[1+(-1)^k e^{-2\beta^2}]}$ and $\left | 0\right \rangle$ denotes the vacuum state.

In this task, we consider five bosonic modes in the QNN and start the system from a coherent state ($\left | \beta \right \rangle$) for mode $1$ and vacuum states ($\left | 0 \right \rangle$) for other modes, see the schematic in Fig.~\ref{Figure2}\textbf{(a)}. Each mode is under a coherent pump and the system evolves following Eq.~\ref{Equation2} until time $\tau$. The reservoir modes can emit, e.g., photons, which can be recombined through linear mixing to form a set of new bosonic modes ($\hat b_i$) satisfying $[\hat b_i,\hat b_j^{\dagger}] = \delta_{ij}$. The linear mixing is energy conserving and can be done with linear optics elements, i.e., beam splitters and phase shifters~\cite{Carolan2015}. The modes after the linear mixing process can be represented by their annihilation (creation) operator $\hat c_i$ ($\hat c_i^{\dagger}$), where $\hat c_i = \sum_{j} \textbf{W}_{ij} \hat b_j$, where $W_{ij}$ is the weight matrix that is unitary, following the commutation conditions from the output modes~\cite{Krisnanda2021}. For a QNN composed of $N$ modes, one can construct at most $N$ output modes with the linear mixing process. In what follows, we will focus on one output mode, while the other $N-1$ are assumed to be vacuum, which can be realized with conditional measurements. These are routinely used to generate Fock states in a strongly coupled oscillator-spin system~\cite{Cirac1993}, motional Fock states of an atom ~\cite{Matos1996}, superposition of pure states \cite{Parkins1993,Song1990,Ogawa1991}, photon added states~\cite{Dakna1999}, etc. Conditional measurement is also demonstrated to have improvement on teleportation of continuous variables ~\cite{Opatrny2000,Cochrane2002,Olivares2003}. Related experiments based on beam splitters and two photon-number-resolving detectors~\cite{Allevi2010} and silicon photomultipliers~\cite{Chesi2021} have also been realized. We note that apart from linear optics, the linear mixing transformation can be obtained with simple tunable hopping interactions between bosonic modes. An exemplary experimental setup in this case has been demonstrated with two interacting microwave cavities~\cite{Gao2018}.

Since the task is generating a corresponding Schr\"{o}dinger cat state, we define the Wigner function of the obtained quantum states from the QNN as $W^O(x,p)$ and the target Schr\"{o}dinger cat state as $W^T(x,p)$, the error can be defined as:

\begin{align}
\delta =\frac{ \int (W^O(x,p) - W^T(x,p))^2 dxdp}{ \int (W^O(x,p) + W^T(x,p))^2 dxdp}.
\label{Equation5}
\end{align}
In this case, the incoherent loss in the QNN plays a role similar to noise in the classical task, i.e., a process that reduces the quality of the output. In our simulation, we first start with different specific coherent states, for example, with amplitude $\beta$ equals $1.4, 1.3, 1.2, 1.1$ and $1$. The aim here is to generate corresponding Schr\"{o}dinger cat states from different initial states $|\beta\rangle$ by optimizing the connection weights from the QNN to the new constructed modes. Fig.~\ref{Figure2}\textbf{(c)} presents a prepared Schr\"{o}dinger cat state with an error $\delta=0.07$, while the conditional measurement probability is $0.028$. One can compare it with Fig.~\ref{Figure2}\textbf{(b)} to see that it is well matched with the target state. In Fig.~\ref{Figure3}, we can also see that with increasing strength of Kerr nonlinearity ($\alpha$), the error decreases down to $\sim 0.03$ for $\beta=1$, which proves that the Kerr nonlinearity is helping the QNN functioning in a noisy environment. We also consider when the strength of Kerr nonlinearity goes to infinity, in which case the bosonic modes are effective fermions. Fig.~\ref{Figure3} shows that with infinite Kerr nonlinearity, the QNN still offers low enough error $\delta$ to generate the Schr\"{o}dinger cat states. The conditional measurement probability for infinite Kerr nonlinearity (fermions) to achieve comparable error $\delta$ to the results in Fig.~\ref{Figure2}\textbf{(c)} is $0.085$. Additionally, we note that with decreasing values of $\beta$, the error decreases as well, which indicates that for smaller amplitude $\beta$, the corresponding Schr\"{o}dinger cat states can be easier to obtain.

In this process, we keep the nearest neighbour coupling $\{ J_{ij} \}$ random and fixed. Moreover, for a general assessment, we generate $10$ random coherent states with $\beta\in [1,1.4]$ and use the same optimisation scheme as described before. The average error under different strengths of nonlinearity is presented in Fig.~\ref{Figure3}\textbf{(b)}, which shows similar trend to that in Fig.~\ref{Figure3}\textbf{(a)}. Our results show that in general, i.e., for different coherent states as input, the Kerr nonlinearity allows error resistant production of Schr\"{o}dinger cat states. We have also considered other quantum tasks, such as single photon state generation, and found the same conclusion (see SM).
\begin{figure}[h!]
\includegraphics[width=1\columnwidth]{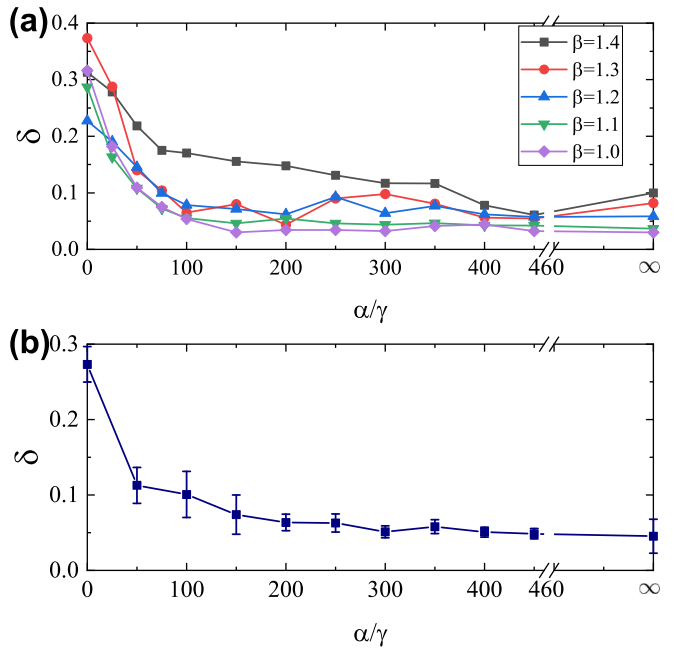}
\caption{\textbf{(a)} The error $E$ as a function of different strength of Kerr nonlinearity $\alpha$. Different color indicates different amplitudes of the input coherent states $\left | \beta \right \rangle$.
\textbf{(b)} Averaged error $E$ as a function of the strength of Kerr nonlinearity $\alpha$ when the coherent states are generated randomly ($\beta$ is randomly distributed between $1.0$ and $1.4$). Each data point is averaged by $10$ times.}
\label{Figure4}
\end{figure}

\emph{Conclusion.} We demonstrated how the Kerr nonlinearity functions in quantum machine learning with a random nearest neighbour coupling bosonic QNN by considering both classical and quantum tasks. Starting with a classical task, we simulate the XOR gate. We show that the nonlinear input-output mapping is able to perform nonlinear classical tasks, for example, XOR gate, even without the Kerr nonlinearity. In a practical environment, the measurement error is unavoidable. We then introduce error on the measured occupation number. The Kerr nonlinearity is shown to be able to correct the error. When considering the quantum task, we construct a quantum gate operating on coherent states and generating corresponding Schr\"{o}dinger cat states. The incoherent loss is interpreted as error in this system, the results show that Kerr nonlinearity offers the capability to resist the error/noise. These results give a clear direction of how to construct and optimize a QNN for performing different tasks (both in classical and quantum regime).
\vspace{0.5cm}

\emph{Acknowledgement.}
This work was supported by the Singapore Ministry of Education under its AcRF Tier 2 grant MOE2019-T2-1-004.

\bibliography{references}

\begin{thebibliography}{59}%
\makeatletter
\providecommand \@ifxundefined [1]{%
 \@ifx{#1\undefined}
}%
\providecommand \@ifnum [1]{%
 \ifnum #1\expandafter \@firstoftwo
 \else \expandafter \@secondoftwo
 \fi
}%
\providecommand \@ifx [1]{%
 \ifx #1\expandafter \@firstoftwo
 \else \expandafter \@secondoftwo
 \fi
}%
\providecommand \natexlab [1]{#1}%
\providecommand \enquote  [1]{``#1''}%
\providecommand \bibnamefont  [1]{#1}%
\providecommand \bibfnamefont [1]{#1}%
\providecommand \citenamefont [1]{#1}%
\providecommand \href@noop [0]{\@secondoftwo}%
\providecommand \href [0]{\begingroup \@sanitize@url \@href}%
\providecommand \@href[1]{\@@startlink{#1}\@@href}%
\providecommand \@@href[1]{\endgroup#1\@@endlink}%
\providecommand \@sanitize@url [0]{\catcode `\\12\catcode `\$12\catcode
  `\&12\catcode `\#12\catcode `\^12\catcode `\_12\catcode `\%12\relax}%
\providecommand \@@startlink[1]{}%
\providecommand \@@endlink[0]{}%
\providecommand \url  [0]{\begingroup\@sanitize@url \@url }%
\providecommand \@url [1]{\endgroup\@href {#1}{\urlprefix }}%
\providecommand \urlprefix  [0]{URL }%
\providecommand \Eprint [0]{\href }%
\providecommand \doibase [0]{http://dx.doi.org/}%
\providecommand \selectlanguage [0]{\@gobble}%
\providecommand \bibinfo  [0]{\@secondoftwo}%
\providecommand \bibfield  [0]{\@secondoftwo}%
\providecommand \translation [1]{[#1]}%
\providecommand \BibitemOpen [0]{}%
\providecommand \bibitemStop [0]{}%
\providecommand \bibitemNoStop [0]{.\EOS\space}%
\providecommand \EOS [0]{\spacefactor3000\relax}%
\providecommand \BibitemShut  [1]{\csname bibitem#1\endcsname}%
\let\auto@bib@innerbib\@empty
\bibitem [{\citenamefont {Jelena}\ \emph {et~al.}(2015)\citenamefont {Jelena},
  \citenamefont {Richard}, \citenamefont {Gilbert},\ and\ \citenamefont
  {Brad}}]{Jelena2015}%
  \BibitemOpen
  \bibfield  {author} {\bibinfo {author} {\bibfnamefont {S.}~\bibnamefont
  {Jelena}}, \bibinfo {author} {\bibfnamefont {S.}~\bibnamefont {Richard}},
  \bibinfo {author} {\bibfnamefont {C.}~\bibnamefont {Gilbert}}, \ and\
  \bibinfo {author} {\bibfnamefont {W.}~\bibnamefont {Brad}},\ }\href {\doibase
  10.1126/science.349.6245.248} {\bibfield  {journal} {\bibinfo  {journal}
  {Science}\ }\textbf {\bibinfo {volume} {349}},\ \bibinfo {pages} {248}
  (\bibinfo {year} {2015})}\BibitemShut {NoStop}%
\bibitem [{\citenamefont {Chouard}\ and\ \citenamefont
  {Venema}(2015)}]{Chouard2015}%
  \BibitemOpen
  \bibfield  {author} {\bibinfo {author} {\bibfnamefont {T.}~\bibnamefont
  {Chouard}}\ and\ \bibinfo {author} {\bibfnamefont {L.}~\bibnamefont
  {Venema}},\ }\href {\doibase 10.1038/521435a} {\bibfield  {journal} {\bibinfo
   {journal} {Nature}\ }\textbf {\bibinfo {volume} {521}},\ \bibinfo {pages}
  {435} (\bibinfo {year} {2015})}\BibitemShut {NoStop}%
\bibitem [{\citenamefont {LeCun}\ \emph {et~al.}(2015)\citenamefont {LeCun},
  \citenamefont {Bengio},\ and\ \citenamefont {Hinton}}]{LeCun2015}%
  \BibitemOpen
  \bibfield  {author} {\bibinfo {author} {\bibfnamefont {Y.}~\bibnamefont
  {LeCun}}, \bibinfo {author} {\bibfnamefont {Y.}~\bibnamefont {Bengio}}, \
  and\ \bibinfo {author} {\bibfnamefont {G.}~\bibnamefont {Hinton}},\ }\href
  {\doibase 10.1038/nature14539} {\bibfield  {journal} {\bibinfo  {journal}
  {Nature}\ }\textbf {\bibinfo {volume} {521}},\ \bibinfo {pages} {436}
  (\bibinfo {year} {2015})}\BibitemShut {NoStop}%
\bibitem [{\citenamefont {Butler}\ \emph {et~al.}(2018)\citenamefont {Butler},
  \citenamefont {Davies}, \citenamefont {Cartwright}, \citenamefont {Isayev},\
  and\ \citenamefont {Walsh}}]{Butler2018}%
  \BibitemOpen
  \bibfield  {author} {\bibinfo {author} {\bibfnamefont {K.~T.}\ \bibnamefont
  {Butler}}, \bibinfo {author} {\bibfnamefont {D.~W.}\ \bibnamefont {Davies}},
  \bibinfo {author} {\bibfnamefont {H.}~\bibnamefont {Cartwright}}, \bibinfo
  {author} {\bibfnamefont {O.}~\bibnamefont {Isayev}}, \ and\ \bibinfo {author}
  {\bibfnamefont {A.}~\bibnamefont {Walsh}},\ }\href {\doibase
  10.1038/s41586-018-0337-2} {\bibfield  {journal} {\bibinfo  {journal}
  {Nature}\ }\textbf {\bibinfo {volume} {559}},\ \bibinfo {pages} {547}
  (\bibinfo {year} {2018})}\BibitemShut {NoStop}%
\bibitem [{\citenamefont {Bebis}\ and\ \citenamefont
  {Georgiopoulos}(1994)}]{Bebis1994}%
  \BibitemOpen
  \bibfield  {author} {\bibinfo {author} {\bibfnamefont {G.}~\bibnamefont
  {Bebis}}\ and\ \bibinfo {author} {\bibfnamefont {M.}~\bibnamefont
  {Georgiopoulos}},\ }\href@noop {} {\bibfield  {journal} {\bibinfo  {journal}
  {IEEE Potentials}\ }\textbf {\bibinfo {volume} {13}},\ \bibinfo {pages} {27}
  (\bibinfo {year} {1994})}\BibitemShut {NoStop}%
\bibitem [{\citenamefont {Herbert}\ and\ \citenamefont
  {Harald}(2004)}]{Herbert2004}%
  \BibitemOpen
  \bibfield  {author} {\bibinfo {author} {\bibfnamefont {J.}~\bibnamefont
  {Herbert}}\ and\ \bibinfo {author} {\bibfnamefont {H.}~\bibnamefont
  {Harald}},\ }\href {\doibase 10.1126/science.1091277} {\bibfield  {journal}
  {\bibinfo  {journal} {Science}\ }\textbf {\bibinfo {volume} {304}},\ \bibinfo
  {pages} {78} (\bibinfo {year} {2004})}\BibitemShut {NoStop}%
\bibitem [{\citenamefont {Maass}\ \emph {et~al.}(2002)\citenamefont {Maass},
  \citenamefont {Natschl{\"a}ger},\ and\ \citenamefont {Markram}}]{Maass2002}%
  \BibitemOpen
  \bibfield  {author} {\bibinfo {author} {\bibfnamefont {W.}~\bibnamefont
  {Maass}}, \bibinfo {author} {\bibfnamefont {T.}~\bibnamefont
  {Natschl{\"a}ger}}, \ and\ \bibinfo {author} {\bibfnamefont {H.}~\bibnamefont
  {Markram}},\ }\href {\doibase 10.1162/089976602760407955} {\bibfield
  {journal} {\bibinfo  {journal} {Neural Computation}\ }\textbf {\bibinfo
  {volume} {14}},\ \bibinfo {pages} {2531} (\bibinfo {year}
  {2002})}\BibitemShut {NoStop}%
\bibitem [{\citenamefont {Enel}\ \emph {et~al.}(2016)\citenamefont {Enel},
  \citenamefont {Procyk}, \citenamefont {Quilodran},\ and\ \citenamefont
  {Dominey}}]{Enel2016}%
  \BibitemOpen
  \bibfield  {author} {\bibinfo {author} {\bibfnamefont {P.}~\bibnamefont
  {Enel}}, \bibinfo {author} {\bibfnamefont {E.}~\bibnamefont {Procyk}},
  \bibinfo {author} {\bibfnamefont {R.}~\bibnamefont {Quilodran}}, \ and\
  \bibinfo {author} {\bibfnamefont {P.~F.}\ \bibnamefont {Dominey}},\ }\href
  {https://doi.org/10.1371/journal.pcbi.1004967} {\bibfield  {journal}
  {\bibinfo  {journal} {PLOS Computational Biology}\ }\textbf {\bibinfo
  {volume} {12}},\ \bibinfo {pages} {e1004967} (\bibinfo {year}
  {2016})}\BibitemShut {NoStop}%
\bibitem [{\citenamefont {Lukosevicius}\ and\ \citenamefont
  {Jaeger}(2009)}]{Lukosevicius2009}%
  \BibitemOpen
  \bibfield  {author} {\bibinfo {author} {\bibfnamefont {M.}~\bibnamefont
  {Lukosevicius}}\ and\ \bibinfo {author} {\bibfnamefont {H.}~\bibnamefont
  {Jaeger}},\ }\href {\doibase https://doi.org/10.1016/j.cosrev.2009.03.005}
  {\bibfield  {journal} {\bibinfo  {journal} {Computer Science Review}\
  }\textbf {\bibinfo {volume} {3}},\ \bibinfo {pages} {127} (\bibinfo {year}
  {2009})}\BibitemShut {NoStop}%
\bibitem [{\citenamefont {Deng}\ \emph {et~al.}(2013)\citenamefont {Deng},
  \citenamefont {Hinton},\ and\ \citenamefont {Kingsbury}}]{Deng2013}%
  \BibitemOpen
  \bibfield  {author} {\bibinfo {author} {\bibfnamefont {L.}~\bibnamefont
  {Deng}}, \bibinfo {author} {\bibfnamefont {G.}~\bibnamefont {Hinton}}, \ and\
  \bibinfo {author} {\bibfnamefont {B.}~\bibnamefont {Kingsbury}},\ }in\ \href
  {\doibase 10.1109/ICASSP.2013.6639344} {\emph {\bibinfo {booktitle} {2013
  IEEE International Conference on Acoustics, Speech and Signal Processing}}}\
  (\bibinfo {year} {2013})\ pp.\ \bibinfo {pages} {8599--8603}\BibitemShut
  {NoStop}%
\bibitem [{\citenamefont {Chan}\ \emph {et~al.}(2016)\citenamefont {Chan},
  \citenamefont {Jaitly}, \citenamefont {Le},\ and\ \citenamefont
  {Vinyals}}]{Chan2016}%
  \BibitemOpen
  \bibfield  {author} {\bibinfo {author} {\bibfnamefont {W.}~\bibnamefont
  {Chan}}, \bibinfo {author} {\bibfnamefont {N.}~\bibnamefont {Jaitly}},
  \bibinfo {author} {\bibfnamefont {Q.}~\bibnamefont {Le}}, \ and\ \bibinfo
  {author} {\bibfnamefont {O.}~\bibnamefont {Vinyals}},\ }in\ \href {\doibase
  10.1109/ICASSP.2016.7472621} {\emph {\bibinfo {booktitle} {2016 IEEE
  International Conference on Acoustics, Speech and Signal Processing
  (ICASSP)}}}\ (\bibinfo {year} {2016})\ pp.\ \bibinfo {pages}
  {4960--4964}\BibitemShut {NoStop}%
\bibitem [{\citenamefont {Ciresan}\ \emph {et~al.}(2011)\citenamefont
  {Ciresan}, \citenamefont {Meier}, \citenamefont {Gambardella},\ and\
  \citenamefont {Schmidhuber}}]{Ciresan2011}%
  \BibitemOpen
  \bibfield  {author} {\bibinfo {author} {\bibfnamefont {D.~C.}\ \bibnamefont
  {Ciresan}}, \bibinfo {author} {\bibfnamefont {U.}~\bibnamefont {Meier}},
  \bibinfo {author} {\bibfnamefont {L.~M.}\ \bibnamefont {Gambardella}}, \ and\
  \bibinfo {author} {\bibfnamefont {J.}~\bibnamefont {Schmidhuber}},\ }in\
  \href {\doibase 10.1109/ICDAR.2011.229} {\emph {\bibinfo {booktitle} {2011
  International Conference on Document Analysis and Recognition}}}\ (\bibinfo
  {year} {2011})\ pp.\ \bibinfo {pages} {1135--1139}\BibitemShut {NoStop}%
\bibitem [{\citenamefont {Leung}\ \emph {et~al.}(1991)\citenamefont {Leung},
  \citenamefont {Leung}, \citenamefont {Lau},\ and\ \citenamefont
  {Luk}}]{Leung1991}%
  \BibitemOpen
  \bibfield  {author} {\bibinfo {author} {\bibfnamefont {W.~F.}\ \bibnamefont
  {Leung}}, \bibinfo {author} {\bibfnamefont {S.~H.}\ \bibnamefont {Leung}},
  \bibinfo {author} {\bibfnamefont {W.~H.}\ \bibnamefont {Lau}}, \ and\
  \bibinfo {author} {\bibfnamefont {A.}~\bibnamefont {Luk}},\ }in\ \href
  {\doibase 10.1109/NNSP.1991.239519} {\emph {\bibinfo {booktitle} {Neural
  Networks for Signal Processing Proceedings of the 1991 IEEE Workshop}}}\
  (\bibinfo {year} {1991})\ pp.\ \bibinfo {pages} {226--235}\BibitemShut
  {NoStop}%
\bibitem [{\citenamefont {Kaastra}\ and\ \citenamefont
  {Boyd}(1996)}]{Kaastra1996}%
  \BibitemOpen
  \bibfield  {author} {\bibinfo {author} {\bibfnamefont {I.}~\bibnamefont
  {Kaastra}}\ and\ \bibinfo {author} {\bibfnamefont {M.}~\bibnamefont {Boyd}},\
  }\href {\doibase https://doi.org/10.1016/0925-2312(95)00039-9} {\bibfield
  {journal} {\bibinfo  {journal} {Neurocomputing}\ }\textbf {\bibinfo {volume}
  {10}},\ \bibinfo {pages} {215} (\bibinfo {year} {1996})}\BibitemShut
  {NoStop}%
\bibitem [{\citenamefont {Zhang}\ \emph {et~al.}(2001)\citenamefont {Zhang},
  \citenamefont {Patuwo},\ and\ \citenamefont {Hu}}]{Zhang2001}%
  \BibitemOpen
  \bibfield  {author} {\bibinfo {author} {\bibfnamefont {G.~P.}\ \bibnamefont
  {Zhang}}, \bibinfo {author} {\bibfnamefont {B.~E.}\ \bibnamefont {Patuwo}}, \
  and\ \bibinfo {author} {\bibfnamefont {M.~Y.}\ \bibnamefont {Hu}},\ }\href
  {\doibase https://doi.org/10.1016/S0305-0548(99)00123-9} {\bibfield
  {journal} {\bibinfo  {journal} {Computers \& Operations Research}\ }\textbf
  {\bibinfo {volume} {28}},\ \bibinfo {pages} {381} (\bibinfo {year}
  {2001})}\BibitemShut {NoStop}%
\bibitem [{\citenamefont {Xu}(2017)}]{Xu2017}%
  \BibitemOpen
  \bibfield  {author} {\bibinfo {author} {\bibfnamefont {G.}~\bibnamefont
  {Xu}},\ }in\ \href {\doibase 10.1145/3082031.3083236} {\emph {\bibinfo
  {booktitle} {Proceedings of the 5th ACM Workshop on Information Hiding and
  Multimedia Security}}}\ (\bibinfo  {publisher} {Association for Computing
  Machinery},\ \bibinfo {address} {New York, NY, USA},\ \bibinfo {year}
  {2017})\ pp.\ \bibinfo {pages} {67--73}\BibitemShut {NoStop}%
\bibitem [{\citenamefont {{Liu}}\ \emph {et~al.}(2016)\citenamefont {{Liu}},
  \citenamefont {{Qiu}},\ and\ \citenamefont {{Huang}}}]{Liu2016}%
  \BibitemOpen
  \bibfield  {author} {\bibinfo {author} {\bibfnamefont {P.}~\bibnamefont
  {{Liu}}}, \bibinfo {author} {\bibfnamefont {X.}~\bibnamefont {{Qiu}}}, \ and\
  \bibinfo {author} {\bibfnamefont {X.}~\bibnamefont {{Huang}}},\ }\href@noop
  {} {\bibfield  {journal} {\bibinfo  {journal} {arXiv e-prints}\ ,\ \bibinfo
  {eid} {arXiv:1605.05101}} (\bibinfo {year} {2016})},\ \Eprint
  {http://arxiv.org/abs/1605.05101} {arXiv:1605.05101 [cs.CL]} \BibitemShut
  {NoStop}%
\bibitem [{\citenamefont {Zhang}\ \emph {et~al.}(2019)\citenamefont {Zhang},
  \citenamefont {Zhang}, \citenamefont {Chen}, \citenamefont {Sun},
  \citenamefont {Ma},\ and\ \citenamefont {Yu}}]{Zhang2019}%
  \BibitemOpen
  \bibfield  {author} {\bibinfo {author} {\bibfnamefont {Q.}~\bibnamefont
  {Zhang}}, \bibinfo {author} {\bibfnamefont {M.}~\bibnamefont {Zhang}},
  \bibinfo {author} {\bibfnamefont {T.}~\bibnamefont {Chen}}, \bibinfo {author}
  {\bibfnamefont {Z.}~\bibnamefont {Sun}}, \bibinfo {author} {\bibfnamefont
  {Y.}~\bibnamefont {Ma}}, \ and\ \bibinfo {author} {\bibfnamefont
  {B.}~\bibnamefont {Yu}},\ }\href {\doibase
  https://doi.org/10.1016/j.neucom.2018.09.038} {\bibfield  {journal} {\bibinfo
   {journal} {Neurocomputing}\ }\textbf {\bibinfo {volume} {323}},\ \bibinfo
  {pages} {37} (\bibinfo {year} {2019})}\BibitemShut {NoStop}%
\bibitem [{\citenamefont {Andrew L.~Maas}(2013)}]{Maas2013}%
  \BibitemOpen
  \bibfield  {author} {\bibinfo {author} {\bibfnamefont {A.~Y.~N.}\
  \bibnamefont {Andrew L.~Maas}, \bibfnamefont {Awni Y.~Hannun}},\ }\href@noop
  {} {\bibfield  {journal} {\bibinfo  {journal} {Proc.icml}\ }\textbf {\bibinfo
  {volume} {30}},\ \bibinfo {pages} {3} (\bibinfo {year} {2013})}\BibitemShut
  {NoStop}%
\bibitem [{\citenamefont {Krzysteczko}\ \emph {et~al.}(2012)\citenamefont
  {Krzysteczko}, \citenamefont {M{\"u}nchenberger}, \citenamefont
  {Sch{\"a}fers}, \citenamefont {Reiss},\ and\ \citenamefont
  {Thomas}}]{Krzysteczko2012}%
  \BibitemOpen
  \bibfield  {author} {\bibinfo {author} {\bibfnamefont {P.}~\bibnamefont
  {Krzysteczko}}, \bibinfo {author} {\bibfnamefont {J.}~\bibnamefont
  {M{\"u}nchenberger}}, \bibinfo {author} {\bibfnamefont {M.}~\bibnamefont
  {Sch{\"a}fers}}, \bibinfo {author} {\bibfnamefont {G.}~\bibnamefont {Reiss}},
  \ and\ \bibinfo {author} {\bibfnamefont {A.}~\bibnamefont {Thomas}},\ }\href
  {\doibase https://doi.org/10.1002/adma.201103723} {\bibfield  {journal}
  {\bibinfo  {journal} {Adv. Mater.}\ }\textbf {\bibinfo {volume} {24}},\
  \bibinfo {pages} {762} (\bibinfo {year} {2012})}\BibitemShut {NoStop}%
\bibitem [{\citenamefont {Locatelli}\ \emph {et~al.}(2014)\citenamefont
  {Locatelli}, \citenamefont {Cros},\ and\ \citenamefont
  {Grollier}}]{Locatelli2014}%
  \BibitemOpen
  \bibfield  {author} {\bibinfo {author} {\bibfnamefont {N.}~\bibnamefont
  {Locatelli}}, \bibinfo {author} {\bibfnamefont {V.}~\bibnamefont {Cros}}, \
  and\ \bibinfo {author} {\bibfnamefont {J.}~\bibnamefont {Grollier}},\ }\href
  {\doibase 10.1038/nmat3823} {\bibfield  {journal} {\bibinfo  {journal} {Nat.
  Mater.}\ }\textbf {\bibinfo {volume} {13}},\ \bibinfo {pages} {11} (\bibinfo
  {year} {2014})}\BibitemShut {NoStop}%
\bibitem [{\citenamefont {Quang~Diep}\ \emph {et~al.}(2014)\citenamefont
  {Quang~Diep}, \citenamefont {Sutton}, \citenamefont {Behin-Aein},\ and\
  \citenamefont {Datta}}]{Quang-Diep2014}%
  \BibitemOpen
  \bibfield  {author} {\bibinfo {author} {\bibfnamefont {V.}~\bibnamefont
  {Quang~Diep}}, \bibinfo {author} {\bibfnamefont {B.}~\bibnamefont {Sutton}},
  \bibinfo {author} {\bibfnamefont {B.}~\bibnamefont {Behin-Aein}}, \ and\
  \bibinfo {author} {\bibfnamefont {S.}~\bibnamefont {Datta}},\ }\href
  {\doibase 10.1063/1.4881575} {\bibfield  {journal} {\bibinfo  {journal}
  {Appl. Phys. Lett.}\ }\textbf {\bibinfo {volume} {104}},\ \bibinfo {pages}
  {222405} (\bibinfo {year} {2014})}\BibitemShut {NoStop}%
\bibitem [{\citenamefont {Xu}\ \emph {et~al.}(2020)\citenamefont {Xu},
  \citenamefont {Ghosh}, \citenamefont {Matuszewski},\ and\ \citenamefont
  {Liew}}]{Xu2020}%
  \BibitemOpen
  \bibfield  {author} {\bibinfo {author} {\bibfnamefont {H.}~\bibnamefont
  {Xu}}, \bibinfo {author} {\bibfnamefont {S.}~\bibnamefont {Ghosh}}, \bibinfo
  {author} {\bibfnamefont {M.}~\bibnamefont {Matuszewski}}, \ and\ \bibinfo
  {author} {\bibfnamefont {T.~C.~H.}\ \bibnamefont {Liew}},\ }\href {\doibase
  10.1103/PhysRevApplied.13.064074} {\bibfield  {journal} {\bibinfo  {journal}
  {Phys. Rev. Applied}\ }\textbf {\bibinfo {volume} {13}},\ \bibinfo {pages}
  {064074} (\bibinfo {year} {2020})}\BibitemShut {NoStop}%
\bibitem [{\citenamefont {Opala}\ \emph {et~al.}(2019)\citenamefont {Opala},
  \citenamefont {Ghosh}, \citenamefont {Liew},\ and\ \citenamefont
  {Matuszewski}}]{Opala2019}%
  \BibitemOpen
  \bibfield  {author} {\bibinfo {author} {\bibfnamefont {A.}~\bibnamefont
  {Opala}}, \bibinfo {author} {\bibfnamefont {S.}~\bibnamefont {Ghosh}},
  \bibinfo {author} {\bibfnamefont {T.~C.~H.}\ \bibnamefont {Liew}}, \ and\
  \bibinfo {author} {\bibfnamefont {M.}~\bibnamefont {Matuszewski}},\ }\href
  {\doibase 10.1103/PhysRevApplied.11.064029} {\bibfield  {journal} {\bibinfo
  {journal} {Phys. Rev. Applied}\ }\textbf {\bibinfo {volume} {11}},\ \bibinfo
  {pages} {064029} (\bibinfo {year} {2019})}\BibitemShut {NoStop}%
\bibitem [{\citenamefont {Ballarini}\ \emph {et~al.}(2020)\citenamefont
  {Ballarini}, \citenamefont {Gianfrate}, \citenamefont {Panico}, \citenamefont
  {Opala}, \citenamefont {Ghosh}, \citenamefont {Dominici}, \citenamefont
  {Ardizzone}, \citenamefont {De~Giorgi}, \citenamefont {Lerario},
  \citenamefont {Gigli}, \citenamefont {Liew}, \citenamefont {Matuszewski},\
  and\ \citenamefont {Sanvitto}}]{Ballarini2020}%
  \BibitemOpen
  \bibfield  {author} {\bibinfo {author} {\bibfnamefont {D.}~\bibnamefont
  {Ballarini}}, \bibinfo {author} {\bibfnamefont {A.}~\bibnamefont
  {Gianfrate}}, \bibinfo {author} {\bibfnamefont {R.}~\bibnamefont {Panico}},
  \bibinfo {author} {\bibfnamefont {A.}~\bibnamefont {Opala}}, \bibinfo
  {author} {\bibfnamefont {S.}~\bibnamefont {Ghosh}}, \bibinfo {author}
  {\bibfnamefont {L.}~\bibnamefont {Dominici}}, \bibinfo {author}
  {\bibfnamefont {V.}~\bibnamefont {Ardizzone}}, \bibinfo {author}
  {\bibfnamefont {M.}~\bibnamefont {De~Giorgi}}, \bibinfo {author}
  {\bibfnamefont {G.}~\bibnamefont {Lerario}}, \bibinfo {author} {\bibfnamefont
  {G.}~\bibnamefont {Gigli}}, \bibinfo {author} {\bibfnamefont {T.~C.~H.}\
  \bibnamefont {Liew}}, \bibinfo {author} {\bibfnamefont {M.}~\bibnamefont
  {Matuszewski}}, \ and\ \bibinfo {author} {\bibfnamefont {D.}~\bibnamefont
  {Sanvitto}},\ }\href {\doibase 10.1021/acs.nanolett.0c00435} {\bibfield
  {journal} {\bibinfo  {journal} {Nano Lett.}\ }\textbf {\bibinfo {volume}
  {20}},\ \bibinfo {pages} {3506} (\bibinfo {year} {2020})}\BibitemShut
  {NoStop}%
\bibitem [{\citenamefont {Matuszewski}\ \emph {et~al.}(2021)\citenamefont
  {Matuszewski}, \citenamefont {Opala}, \citenamefont {Mirek}, \citenamefont
  {Furman}, \citenamefont {Kr{\'o}l}, \citenamefont {Tyszka}, \citenamefont
  {Liew}, \citenamefont {Ballarini}, \citenamefont {Sanvitto}, \citenamefont
  {Szczytko},\ and\ \citenamefont {Pi{\k e}tka}}]{Matuszewski2021}%
  \BibitemOpen
  \bibfield  {author} {\bibinfo {author} {\bibfnamefont {M.}~\bibnamefont
  {Matuszewski}}, \bibinfo {author} {\bibfnamefont {A.}~\bibnamefont {Opala}},
  \bibinfo {author} {\bibfnamefont {R.}~\bibnamefont {Mirek}}, \bibinfo
  {author} {\bibfnamefont {M.}~\bibnamefont {Furman}}, \bibinfo {author}
  {\bibfnamefont {M.}~\bibnamefont {Kr{\'o}l}}, \bibinfo {author}
  {\bibfnamefont {K.}~\bibnamefont {Tyszka}}, \bibinfo {author} {\bibfnamefont
  {T.~C.~H.}\ \bibnamefont {Liew}}, \bibinfo {author} {\bibfnamefont
  {D.}~\bibnamefont {Ballarini}}, \bibinfo {author} {\bibfnamefont
  {D.}~\bibnamefont {Sanvitto}}, \bibinfo {author} {\bibfnamefont
  {J.}~\bibnamefont {Szczytko}}, \ and\ \bibinfo {author} {\bibfnamefont
  {B.}~\bibnamefont {Pi{\k e}tka}},\ }\href {\doibase
  10.1103/PhysRevApplied.16.024045} {\bibfield  {journal} {\bibinfo  {journal}
  {Phys. Rev. Applied}\ }\textbf {\bibinfo {volume} {16}},\ \bibinfo {pages}
  {024045} (\bibinfo {year} {2021})}\BibitemShut {NoStop}%
\bibitem [{\citenamefont {Biamonte}\ \emph {et~al.}(2017)\citenamefont
  {Biamonte}, \citenamefont {Wittek}, \citenamefont {Pancotti}, \citenamefont
  {Rebentrost}, \citenamefont {Wiebe},\ and\ \citenamefont
  {Lloyd}}]{Biamonte2017}%
  \BibitemOpen
  \bibfield  {author} {\bibinfo {author} {\bibfnamefont {J.}~\bibnamefont
  {Biamonte}}, \bibinfo {author} {\bibfnamefont {P.}~\bibnamefont {Wittek}},
  \bibinfo {author} {\bibfnamefont {N.}~\bibnamefont {Pancotti}}, \bibinfo
  {author} {\bibfnamefont {P.}~\bibnamefont {Rebentrost}}, \bibinfo {author}
  {\bibfnamefont {N.}~\bibnamefont {Wiebe}}, \ and\ \bibinfo {author}
  {\bibfnamefont {S.}~\bibnamefont {Lloyd}},\ }\href {\doibase
  10.1038/nature23474} {\bibfield  {journal} {\bibinfo  {journal} {Nature}\
  }\textbf {\bibinfo {volume} {549}},\ \bibinfo {pages} {195} (\bibinfo {year}
  {2017})}\BibitemShut {NoStop}%
\bibitem [{\citenamefont {Dunjko}\ and\ \citenamefont
  {Briegel}(2018)}]{Dunjko2018}%
  \BibitemOpen
  \bibfield  {author} {\bibinfo {author} {\bibfnamefont {V.}~\bibnamefont
  {Dunjko}}\ and\ \bibinfo {author} {\bibfnamefont {H.~J.}\ \bibnamefont
  {Briegel}},\ }\href {\doibase 10.1088/1361-6633/aab406} {\bibfield  {journal}
  {\bibinfo  {journal} {Rep. Prog. Phys.}\ }\textbf {\bibinfo {volume} {81}},\
  \bibinfo {pages} {074001} (\bibinfo {year} {2018})}\BibitemShut {NoStop}%
\bibitem [{\citenamefont {Altaisky}\ \emph {et~al.}(2016)\citenamefont
  {Altaisky}, \citenamefont {Zolnikova}, \citenamefont {Kaputkina},
  \citenamefont {Krylov}, \citenamefont {Lozovik},\ and\ \citenamefont
  {Dattani}}]{Altaisky2016}%
  \BibitemOpen
  \bibfield  {author} {\bibinfo {author} {\bibfnamefont {M.~V.}\ \bibnamefont
  {Altaisky}}, \bibinfo {author} {\bibfnamefont {N.~N.}\ \bibnamefont
  {Zolnikova}}, \bibinfo {author} {\bibfnamefont {N.~E.}\ \bibnamefont
  {Kaputkina}}, \bibinfo {author} {\bibfnamefont {V.~A.}\ \bibnamefont
  {Krylov}}, \bibinfo {author} {\bibfnamefont {Y.~E.}\ \bibnamefont {Lozovik}},
  \ and\ \bibinfo {author} {\bibfnamefont {N.~S.}\ \bibnamefont {Dattani}},\
  }\href {\doibase 10.1063/1.4943622} {\bibfield  {journal} {\bibinfo
  {journal} {Appl. Phys. Lett.}\ }\textbf {\bibinfo {volume} {108}},\ \bibinfo
  {pages} {103108} (\bibinfo {year} {2016})}\BibitemShut {NoStop}%
\bibitem [{\citenamefont {McClean}\ \emph {et~al.}(2018)\citenamefont
  {McClean}, \citenamefont {Boixo}, \citenamefont {Smelyanskiy}, \citenamefont
  {Babbush},\ and\ \citenamefont {Neven}}]{McClean2018}%
  \BibitemOpen
  \bibfield  {author} {\bibinfo {author} {\bibfnamefont {J.~R.}\ \bibnamefont
  {McClean}}, \bibinfo {author} {\bibfnamefont {S.}~\bibnamefont {Boixo}},
  \bibinfo {author} {\bibfnamefont {V.~N.}\ \bibnamefont {Smelyanskiy}},
  \bibinfo {author} {\bibfnamefont {R.}~\bibnamefont {Babbush}}, \ and\
  \bibinfo {author} {\bibfnamefont {H.}~\bibnamefont {Neven}},\ }\href
  {\doibase 10.1038/s41467-018-07090-4} {\bibfield  {journal} {\bibinfo
  {journal} {Nat. Commun.}\ }\textbf {\bibinfo {volume} {9}},\ \bibinfo {pages}
  {4812} (\bibinfo {year} {2018})}\BibitemShut {NoStop}%
\bibitem [{\citenamefont {Shen}\ \emph {et~al.}(2020)\citenamefont {Shen},
  \citenamefont {Zhang}, \citenamefont {You},\ and\ \citenamefont
  {Zhai}}]{Shen2020}%
  \BibitemOpen
  \bibfield  {author} {\bibinfo {author} {\bibfnamefont {H.}~\bibnamefont
  {Shen}}, \bibinfo {author} {\bibfnamefont {P.}~\bibnamefont {Zhang}},
  \bibinfo {author} {\bibfnamefont {Y.-Z.}\ \bibnamefont {You}}, \ and\
  \bibinfo {author} {\bibfnamefont {H.}~\bibnamefont {Zhai}},\ }\href {\doibase
  10.1103/PhysRevLett.124.200504} {\bibfield  {journal} {\bibinfo  {journal}
  {Phys. Rev. Lett.}\ }\textbf {\bibinfo {volume} {124}},\ \bibinfo {pages}
  {200504} (\bibinfo {year} {2020})}\BibitemShut {NoStop}%
\bibitem [{\citenamefont {Szegedy}(2004)}]{Szegedy2004}%
  \BibitemOpen
  \bibfield  {author} {\bibinfo {author} {\bibfnamefont {M.}~\bibnamefont
  {Szegedy}},\ }in\ \href {\doibase 10.1109/FOCS.2004.53} {\emph {\bibinfo
  {booktitle} {45th Annual IEEE Symposium on Foundations of Computer
  Science}}}\ (\bibinfo {year} {2004})\ pp.\ \bibinfo {pages}
  {32--41}\BibitemShut {NoStop}%
\bibitem [{\citenamefont {F.}\ \emph {et~al.}(2014)\citenamefont {F.},
  \citenamefont {Zhihui}, \citenamefont {Joshua}, \citenamefont {Sergio},
  \citenamefont {V.}, \citenamefont {David}, \citenamefont {M.}, \citenamefont
  {A.},\ and\ \citenamefont {Matthias}}]{F2014}%
  \BibitemOpen
  \bibfield  {author} {\bibinfo {author} {\bibfnamefont {R.~T.}\ \bibnamefont
  {F.}}, \bibinfo {author} {\bibfnamefont {W.}~\bibnamefont {Zhihui}}, \bibinfo
  {author} {\bibfnamefont {J.}~\bibnamefont {Joshua}}, \bibinfo {author}
  {\bibfnamefont {B.}~\bibnamefont {Sergio}}, \bibinfo {author} {\bibfnamefont
  {I.~S.}\ \bibnamefont {V.}}, \bibinfo {author} {\bibfnamefont
  {W.}~\bibnamefont {David}}, \bibinfo {author} {\bibfnamefont {M.~J.}\
  \bibnamefont {M.}}, \bibinfo {author} {\bibfnamefont {L.~D.}\ \bibnamefont
  {A.}}, \ and\ \bibinfo {author} {\bibfnamefont {T.}~\bibnamefont
  {Matthias}},\ }\href {\doibase 10.1126/science.1252319} {\bibfield  {journal}
  {\bibinfo  {journal} {Science}\ }\textbf {\bibinfo {volume} {345}},\ \bibinfo
  {pages} {420} (\bibinfo {year} {2014})}\BibitemShut {NoStop}%
\bibitem [{\citenamefont {Dunjko}\ \emph {et~al.}(2016)\citenamefont {Dunjko},
  \citenamefont {Taylor},\ and\ \citenamefont {Briegel}}]{Dunjko2016}%
  \BibitemOpen
  \bibfield  {author} {\bibinfo {author} {\bibfnamefont {V.}~\bibnamefont
  {Dunjko}}, \bibinfo {author} {\bibfnamefont {J.~M.}\ \bibnamefont {Taylor}},
  \ and\ \bibinfo {author} {\bibfnamefont {H.~J.}\ \bibnamefont {Briegel}},\
  }\href {\doibase 10.1103/PhysRevLett.117.130501} {\bibfield  {journal}
  {\bibinfo  {journal} {Phys. Rev. Lett.}\ }\textbf {\bibinfo {volume} {117}},\
  \bibinfo {pages} {130501} (\bibinfo {year} {2016})}\BibitemShut {NoStop}%
\bibitem [{\citenamefont {Paparo}\ \emph {et~al.}(2014)\citenamefont {Paparo},
  \citenamefont {Dunjko}, \citenamefont {Makmal}, \citenamefont
  {Martin-Delgado},\ and\ \citenamefont {Briegel}}]{Paparo2014}%
  \BibitemOpen
  \bibfield  {author} {\bibinfo {author} {\bibfnamefont {G.~D.}\ \bibnamefont
  {Paparo}}, \bibinfo {author} {\bibfnamefont {V.}~\bibnamefont {Dunjko}},
  \bibinfo {author} {\bibfnamefont {A.}~\bibnamefont {Makmal}}, \bibinfo
  {author} {\bibfnamefont {M.~A.}\ \bibnamefont {Martin-Delgado}}, \ and\
  \bibinfo {author} {\bibfnamefont {H.~J.}\ \bibnamefont {Briegel}},\ }\href
  {\doibase 10.1103/PhysRevX.4.031002} {\bibfield  {journal} {\bibinfo
  {journal} {Phys. Rev. X}\ }\textbf {\bibinfo {volume} {4}},\ \bibinfo {pages}
  {031002} (\bibinfo {year} {2014})}\BibitemShut {NoStop}%
\bibitem [{\citenamefont {Xu}\ \emph {et~al.}(2021)\citenamefont {Xu},
  \citenamefont {Krisnanda}, \citenamefont {Verstraelen}, \citenamefont
  {Liew},\ and\ \citenamefont {Ghosh}}]{Xu2021}%
  \BibitemOpen
  \bibfield  {author} {\bibinfo {author} {\bibfnamefont {H.}~\bibnamefont
  {Xu}}, \bibinfo {author} {\bibfnamefont {T.}~\bibnamefont {Krisnanda}},
  \bibinfo {author} {\bibfnamefont {W.}~\bibnamefont {Verstraelen}}, \bibinfo
  {author} {\bibfnamefont {T.~C.~H.}\ \bibnamefont {Liew}}, \ and\ \bibinfo
  {author} {\bibfnamefont {S.}~\bibnamefont {Ghosh}},\ }\href {\doibase
  10.1103/PhysRevB.103.195302} {\bibfield  {journal} {\bibinfo  {journal}
  {Phys. Rev. B}\ }\textbf {\bibinfo {volume} {103}},\ \bibinfo {pages}
  {195302} (\bibinfo {year} {2021})}\BibitemShut {NoStop}%
\bibitem [{\citenamefont {Nakajima}\ \emph {et~al.}(2019)\citenamefont
  {Nakajima}, \citenamefont {Fujii}, \citenamefont {Negoro}, \citenamefont
  {Mitarai},\ and\ \citenamefont {Kitagawa}}]{Nakajima2019}%
  \BibitemOpen
  \bibfield  {author} {\bibinfo {author} {\bibfnamefont {K.}~\bibnamefont
  {Nakajima}}, \bibinfo {author} {\bibfnamefont {K.}~\bibnamefont {Fujii}},
  \bibinfo {author} {\bibfnamefont {M.}~\bibnamefont {Negoro}}, \bibinfo
  {author} {\bibfnamefont {K.}~\bibnamefont {Mitarai}}, \ and\ \bibinfo
  {author} {\bibfnamefont {M.}~\bibnamefont {Kitagawa}},\ }\href {\doibase
  10.1103/PhysRevApplied.11.034021} {\bibfield  {journal} {\bibinfo  {journal}
  {Phys. Rev. Applied}\ }\textbf {\bibinfo {volume} {11}},\ \bibinfo {pages}
  {034021} (\bibinfo {year} {2019})}\BibitemShut {NoStop}%
\bibitem [{\citenamefont {Fujii}\ and\ \citenamefont
  {Nakajima}(2017)}]{Fuji2017}%
  \BibitemOpen
  \bibfield  {author} {\bibinfo {author} {\bibfnamefont {K.}~\bibnamefont
  {Fujii}}\ and\ \bibinfo {author} {\bibfnamefont {K.}~\bibnamefont
  {Nakajima}},\ }\href {\doibase 10.1103/PhysRevApplied.8.024030} {\bibfield
  {journal} {\bibinfo  {journal} {Phys. Rev. Applied}\ }\textbf {\bibinfo
  {volume} {8}},\ \bibinfo {pages} {024030} (\bibinfo {year}
  {2017})}\BibitemShut {NoStop}%
\bibitem [{\citenamefont {Neigovzen}\ \emph {et~al.}(2009)\citenamefont
  {Neigovzen}, \citenamefont {Neves}, \citenamefont {Sollacher},\ and\
  \citenamefont {Glaser}}]{Neigovzen2009}%
  \BibitemOpen
  \bibfield  {author} {\bibinfo {author} {\bibfnamefont {R.}~\bibnamefont
  {Neigovzen}}, \bibinfo {author} {\bibfnamefont {J.~L.}\ \bibnamefont
  {Neves}}, \bibinfo {author} {\bibfnamefont {R.}~\bibnamefont {Sollacher}}, \
  and\ \bibinfo {author} {\bibfnamefont {S.~J.}\ \bibnamefont {Glaser}},\
  }\href {\doibase 10.1103/PhysRevA.79.042321} {\bibfield  {journal} {\bibinfo
  {journal} {Phys. Rev. A}\ }\textbf {\bibinfo {volume} {79}},\ \bibinfo
  {pages} {042321} (\bibinfo {year} {2009})}\BibitemShut {NoStop}%
\bibitem [{\citenamefont {Ghosh}\ \emph
  {et~al.}(2019{\natexlab{a}})\citenamefont {Ghosh}, \citenamefont {Opala},
  \citenamefont {Matuszewski}, \citenamefont {Paterek},\ and\ \citenamefont
  {Liew}}]{Ghosh2019a}%
  \BibitemOpen
  \bibfield  {author} {\bibinfo {author} {\bibfnamefont {S.}~\bibnamefont
  {Ghosh}}, \bibinfo {author} {\bibfnamefont {A.}~\bibnamefont {Opala}},
  \bibinfo {author} {\bibfnamefont {M.}~\bibnamefont {Matuszewski}}, \bibinfo
  {author} {\bibfnamefont {T.}~\bibnamefont {Paterek}}, \ and\ \bibinfo
  {author} {\bibfnamefont {T.~C.~H.}\ \bibnamefont {Liew}},\ }\href {\doibase
  10.1038/s41534-019-0149-8} {\bibfield  {journal} {\bibinfo  {journal} {Npj
  Quantum Inf.}\ }\textbf {\bibinfo {volume} {5}},\ \bibinfo {pages} {35}
  (\bibinfo {year} {2019}{\natexlab{a}})}\BibitemShut {NoStop}%
\bibitem [{\citenamefont {Krisnanda}\ \emph
  {et~al.}(2021{\natexlab{a}})\citenamefont {Krisnanda}, \citenamefont {Ghosh},
  \citenamefont {Paterek}, \citenamefont {Laskowski},\ and\ \citenamefont
  {Liew}}]{krisnanda2021beating}%
  \BibitemOpen
  \bibfield  {author} {\bibinfo {author} {\bibfnamefont {T.}~\bibnamefont
  {Krisnanda}}, \bibinfo {author} {\bibfnamefont {S.}~\bibnamefont {Ghosh}},
  \bibinfo {author} {\bibfnamefont {T.}~\bibnamefont {Paterek}}, \bibinfo
  {author} {\bibfnamefont {W.}~\bibnamefont {Laskowski}}, \ and\ \bibinfo
  {author} {\bibfnamefont {T.~C.~H.}\ \bibnamefont {Liew}},\ }\href@noop {} {}
  (\bibinfo {year} {2021}{\natexlab{a}}),\ \Eprint
  {http://arxiv.org/abs/2110.07507} {arXiv:2110.07507 [quant-ph]} \BibitemShut
  {NoStop}%
\bibitem [{\citenamefont {Krisnanda}\ \emph
  {et~al.}(2021{\natexlab{b}})\citenamefont {Krisnanda}, \citenamefont {Ghosh},
  \citenamefont {Paterek},\ and\ \citenamefont {Liew}}]{Krisnanda2021}%
  \BibitemOpen
  \bibfield  {author} {\bibinfo {author} {\bibfnamefont {T.}~\bibnamefont
  {Krisnanda}}, \bibinfo {author} {\bibfnamefont {S.}~\bibnamefont {Ghosh}},
  \bibinfo {author} {\bibfnamefont {T.}~\bibnamefont {Paterek}}, \ and\
  \bibinfo {author} {\bibfnamefont {T.~C.~H.}\ \bibnamefont {Liew}},\ }\href
  {\doibase https://doi.org/10.1016/j.neunet.2021.01.003} {\bibfield  {journal}
  {\bibinfo  {journal} {Neural Netw.}\ }\textbf {\bibinfo {volume} {136}},\
  \bibinfo {pages} {141} (\bibinfo {year} {2021}{\natexlab{b}})}\BibitemShut
  {NoStop}%
\bibitem [{\citenamefont {Ghosh}\ \emph
  {et~al.}(2019{\natexlab{b}})\citenamefont {Ghosh}, \citenamefont {Paterek},\
  and\ \citenamefont {Liew}}]{Ghosh2019}%
  \BibitemOpen
  \bibfield  {author} {\bibinfo {author} {\bibfnamefont {S.}~\bibnamefont
  {Ghosh}}, \bibinfo {author} {\bibfnamefont {T.}~\bibnamefont {Paterek}}, \
  and\ \bibinfo {author} {\bibfnamefont {T.~C.~H.}\ \bibnamefont {Liew}},\
  }\href {\doibase 10.1103/PhysRevLett.123.260404} {\bibfield  {journal}
  {\bibinfo  {journal} {Phys. Rev. Lett.}\ }\textbf {\bibinfo {volume} {123}},\
  \bibinfo {pages} {260404} (\bibinfo {year} {2019}{\natexlab{b}})}\BibitemShut
  {NoStop}%
\bibitem [{\citenamefont {Ghosh}\ \emph {et~al.}(2021)\citenamefont {Ghosh},
  \citenamefont {Opala}, \citenamefont {Matuszewski}, \citenamefont {Paterek},\
  and\ \citenamefont {Liew}}]{Ghosh2021}%
  \BibitemOpen
  \bibfield  {author} {\bibinfo {author} {\bibfnamefont {S.}~\bibnamefont
  {Ghosh}}, \bibinfo {author} {\bibfnamefont {A.}~\bibnamefont {Opala}},
  \bibinfo {author} {\bibfnamefont {M.}~\bibnamefont {Matuszewski}}, \bibinfo
  {author} {\bibfnamefont {T.}~\bibnamefont {Paterek}}, \ and\ \bibinfo
  {author} {\bibfnamefont {T.~C.~H.}\ \bibnamefont {Liew}},\ }\href {\doibase
  10.1109/TNNLS.2020.3009716} {\bibfield  {journal} {\bibinfo  {journal} {IEEE
  Trans. Neural Netw. Learn. Syst.}\ }\textbf {\bibinfo {volume} {32}},\
  \bibinfo {pages} {3148} (\bibinfo {year} {2021})}\BibitemShut {NoStop}%
\bibitem [{\citenamefont {{Mujal}}\ \emph {et~al.}(2021)\citenamefont
  {{Mujal}}, \citenamefont {{Nokkala}}, \citenamefont
  {{Mart{\'\i}nez-Pe{\~n}a}}, \citenamefont {{Giorgi}}, \citenamefont
  {{Soriano}},\ and\ \citenamefont {{Zambrini}}}]{Mujal2021}%
  \BibitemOpen
  \bibfield  {author} {\bibinfo {author} {\bibfnamefont {P.}~\bibnamefont
  {{Mujal}}}, \bibinfo {author} {\bibfnamefont {J.}~\bibnamefont {{Nokkala}}},
  \bibinfo {author} {\bibfnamefont {R.}~\bibnamefont
  {{Mart{\'\i}nez-Pe{\~n}a}}}, \bibinfo {author} {\bibfnamefont {G.~L.}\
  \bibnamefont {{Giorgi}}}, \bibinfo {author} {\bibfnamefont {M.~C.}\
  \bibnamefont {{Soriano}}}, \ and\ \bibinfo {author} {\bibfnamefont
  {R.}~\bibnamefont {{Zambrini}}},\ }\href {\doibase 10.1088/2632-072X/ac340e}
  {\bibfield  {journal} {\bibinfo  {journal} {Journal of Physics: Complexity}\
  }\textbf {\bibinfo {volume} {2}},\ \bibinfo {eid} {045008} (\bibinfo {year}
  {2021})}\BibitemShut {NoStop}%
\bibitem [{\citenamefont {Mirek}\ \emph {et~al.}(2021)\citenamefont {Mirek},
  \citenamefont {Opala}, \citenamefont {Comaron}, \citenamefont {Furman},
  \citenamefont {Kr{\'o}l}, \citenamefont {Tyszka}, \citenamefont
  {Seredy{\'n}ski}, \citenamefont {Ballarini}, \citenamefont {Sanvitto},
  \citenamefont {Liew}, \citenamefont {Pacuski}, \citenamefont
  {Suffczy{\'n}ski}, \citenamefont {Szczytko}, \citenamefont {Matuszewski},\
  and\ \citenamefont {Pi{\k e}tka}}]{Mirek2021}%
  \BibitemOpen
  \bibfield  {author} {\bibinfo {author} {\bibfnamefont {R.}~\bibnamefont
  {Mirek}}, \bibinfo {author} {\bibfnamefont {A.}~\bibnamefont {Opala}},
  \bibinfo {author} {\bibfnamefont {P.}~\bibnamefont {Comaron}}, \bibinfo
  {author} {\bibfnamefont {M.}~\bibnamefont {Furman}}, \bibinfo {author}
  {\bibfnamefont {M.}~\bibnamefont {Kr{\'o}l}}, \bibinfo {author}
  {\bibfnamefont {K.}~\bibnamefont {Tyszka}}, \bibinfo {author} {\bibfnamefont
  {B.}~\bibnamefont {Seredy{\'n}ski}}, \bibinfo {author} {\bibfnamefont
  {D.}~\bibnamefont {Ballarini}}, \bibinfo {author} {\bibfnamefont
  {D.}~\bibnamefont {Sanvitto}}, \bibinfo {author} {\bibfnamefont {T.~C.~H.}\
  \bibnamefont {Liew}}, \bibinfo {author} {\bibfnamefont {W.}~\bibnamefont
  {Pacuski}}, \bibinfo {author} {\bibfnamefont {J.}~\bibnamefont
  {Suffczy{\'n}ski}}, \bibinfo {author} {\bibfnamefont {J.}~\bibnamefont
  {Szczytko}}, \bibinfo {author} {\bibfnamefont {M.}~\bibnamefont
  {Matuszewski}}, \ and\ \bibinfo {author} {\bibfnamefont {B.}~\bibnamefont
  {Pi{\k e}tka}},\ }\href {\doibase 10.1021/acs.nanolett.0c04696} {\bibfield
  {journal} {\bibinfo  {journal} {Nano Letters}\ }\textbf {\bibinfo {volume}
  {21}},\ \bibinfo {pages} {3715} (\bibinfo {year} {2021})}\BibitemShut
  {NoStop}%
\bibitem [{\citenamefont {Carolan}\ \emph {et~al.}(2015)\citenamefont
  {Carolan}, \citenamefont {Harrold}, \citenamefont {Sparrow}, \citenamefont
  {Mart{\'\i}n-L{\'o}pez}, \citenamefont {Russell}, \citenamefont
  {Silverstone}, \citenamefont {Shadbolt}, \citenamefont {Matsuda},
  \citenamefont {Oguma}, \citenamefont {Itoh}, \citenamefont {Marshall},
  \citenamefont {Thompson}, \citenamefont {Matthews}, \citenamefont
  {Hashimoto}, \citenamefont {O'Brien},\ and\ \citenamefont
  {Laing}}]{Carolan2015}%
  \BibitemOpen
  \bibfield  {author} {\bibinfo {author} {\bibfnamefont {J.}~\bibnamefont
  {Carolan}}, \bibinfo {author} {\bibfnamefont {C.}~\bibnamefont {Harrold}},
  \bibinfo {author} {\bibfnamefont {C.}~\bibnamefont {Sparrow}}, \bibinfo
  {author} {\bibfnamefont {E.}~\bibnamefont {Mart{\'\i}n-L{\'o}pez}}, \bibinfo
  {author} {\bibfnamefont {N.~J.}\ \bibnamefont {Russell}}, \bibinfo {author}
  {\bibfnamefont {J.~W.}\ \bibnamefont {Silverstone}}, \bibinfo {author}
  {\bibfnamefont {P.~J.}\ \bibnamefont {Shadbolt}}, \bibinfo {author}
  {\bibfnamefont {N.}~\bibnamefont {Matsuda}}, \bibinfo {author} {\bibfnamefont
  {M.}~\bibnamefont {Oguma}}, \bibinfo {author} {\bibfnamefont
  {M.}~\bibnamefont {Itoh}}, \bibinfo {author} {\bibfnamefont {G.~D.}\
  \bibnamefont {Marshall}}, \bibinfo {author} {\bibfnamefont {M.~G.}\
  \bibnamefont {Thompson}}, \bibinfo {author} {\bibfnamefont {J.~C.~F.}\
  \bibnamefont {Matthews}}, \bibinfo {author} {\bibfnamefont {T.}~\bibnamefont
  {Hashimoto}}, \bibinfo {author} {\bibfnamefont {J.~L.}\ \bibnamefont
  {O'Brien}}, \ and\ \bibinfo {author} {\bibfnamefont {A.}~\bibnamefont
  {Laing}},\ }\href@noop {} {\bibfield  {journal} {\bibinfo  {journal}
  {Science}\ }\textbf {\bibinfo {volume} {349}},\ \bibinfo {pages} {711}
  (\bibinfo {year} {2015})}\BibitemShut {NoStop}%
\bibitem [{\citenamefont {Cirac}\ \emph {et~al.}(1993)\citenamefont {Cirac},
  \citenamefont {Blatt}, \citenamefont {Parkins},\ and\ \citenamefont
  {Zoller}}]{Cirac1993}%
  \BibitemOpen
  \bibfield  {author} {\bibinfo {author} {\bibfnamefont {J.~I.}\ \bibnamefont
  {Cirac}}, \bibinfo {author} {\bibfnamefont {R.}~\bibnamefont {Blatt}},
  \bibinfo {author} {\bibfnamefont {A.~S.}\ \bibnamefont {Parkins}}, \ and\
  \bibinfo {author} {\bibfnamefont {P.}~\bibnamefont {Zoller}},\ }\href
  {\doibase 10.1103/PhysRevLett.70.762} {\bibfield  {journal} {\bibinfo
  {journal} {Phys. Rev. Lett.}\ }\textbf {\bibinfo {volume} {70}},\ \bibinfo
  {pages} {762} (\bibinfo {year} {1993})}\BibitemShut {NoStop}%
\bibitem [{\citenamefont {de~Matos~Filho}\ and\ \citenamefont
  {Vogel}(1996)}]{Matos1996}%
  \BibitemOpen
  \bibfield  {author} {\bibinfo {author} {\bibfnamefont {R.~L.}\ \bibnamefont
  {de~Matos~Filho}}\ and\ \bibinfo {author} {\bibfnamefont {W.}~\bibnamefont
  {Vogel}},\ }\href {\doibase 10.1103/PhysRevLett.76.4520} {\bibfield
  {journal} {\bibinfo  {journal} {Phys. Rev. Lett.}\ }\textbf {\bibinfo
  {volume} {76}},\ \bibinfo {pages} {4520} (\bibinfo {year}
  {1996})}\BibitemShut {NoStop}%
\bibitem [{\citenamefont {Parkins}\ \emph {et~al.}(1993)\citenamefont
  {Parkins}, \citenamefont {Marte}, \citenamefont {Zoller},\ and\ \citenamefont
  {Kimble}}]{Parkins1993}%
  \BibitemOpen
  \bibfield  {author} {\bibinfo {author} {\bibfnamefont {A.~S.}\ \bibnamefont
  {Parkins}}, \bibinfo {author} {\bibfnamefont {P.}~\bibnamefont {Marte}},
  \bibinfo {author} {\bibfnamefont {P.}~\bibnamefont {Zoller}}, \ and\ \bibinfo
  {author} {\bibfnamefont {H.~J.}\ \bibnamefont {Kimble}},\ }\href {\doibase
  10.1103/PhysRevLett.71.3095} {\bibfield  {journal} {\bibinfo  {journal}
  {Phys. Rev. Lett.}\ }\textbf {\bibinfo {volume} {71}},\ \bibinfo {pages}
  {3095} (\bibinfo {year} {1993})}\BibitemShut {NoStop}%
\bibitem [{\citenamefont {Song}\ \emph {et~al.}(1990)\citenamefont {Song},
  \citenamefont {Caves},\ and\ \citenamefont {Yurke}}]{Song1990}%
  \BibitemOpen
  \bibfield  {author} {\bibinfo {author} {\bibfnamefont {S.}~\bibnamefont
  {Song}}, \bibinfo {author} {\bibfnamefont {C.~M.}\ \bibnamefont {Caves}}, \
  and\ \bibinfo {author} {\bibfnamefont {B.}~\bibnamefont {Yurke}},\ }\href
  {\doibase 10.1103/PhysRevA.41.5261} {\bibfield  {journal} {\bibinfo
  {journal} {Phys. Rev. A}\ }\textbf {\bibinfo {volume} {41}},\ \bibinfo
  {pages} {5261} (\bibinfo {year} {1990})}\BibitemShut {NoStop}%
\bibitem [{\citenamefont {Ogawa}\ \emph {et~al.}(1991)\citenamefont {Ogawa},
  \citenamefont {Ueda},\ and\ \citenamefont {Imoto}}]{Ogawa1991}%
  \BibitemOpen
  \bibfield  {author} {\bibinfo {author} {\bibfnamefont {T.}~\bibnamefont
  {Ogawa}}, \bibinfo {author} {\bibfnamefont {M.}~\bibnamefont {Ueda}}, \ and\
  \bibinfo {author} {\bibfnamefont {N.}~\bibnamefont {Imoto}},\ }\href
  {\doibase 10.1103/PhysRevA.43.6458} {\bibfield  {journal} {\bibinfo
  {journal} {Phys. Rev. A}\ }\textbf {\bibinfo {volume} {43}},\ \bibinfo
  {pages} {6458} (\bibinfo {year} {1991})}\BibitemShut {NoStop}%
\bibitem [{\citenamefont {Dakna}\ \emph {et~al.}(1999)\citenamefont {Dakna},
  \citenamefont {Clausen}, \citenamefont {Kn{\"o}ll},\ and\ \citenamefont
  {Welsch}}]{Dakna1999}%
  \BibitemOpen
  \bibfield  {author} {\bibinfo {author} {\bibfnamefont {M.}~\bibnamefont
  {Dakna}}, \bibinfo {author} {\bibfnamefont {J.}~\bibnamefont {Clausen}},
  \bibinfo {author} {\bibfnamefont {L.}~\bibnamefont {Kn{\"o}ll}}, \ and\
  \bibinfo {author} {\bibfnamefont {D.~G.}\ \bibnamefont {Welsch}},\ }\href
  {\doibase 10.1103/PhysRevA.59.1658} {\bibfield  {journal} {\bibinfo
  {journal} {Phys. Rev. A}\ }\textbf {\bibinfo {volume} {59}},\ \bibinfo
  {pages} {1658} (\bibinfo {year} {1999})}\BibitemShut {NoStop}%
\bibitem [{\citenamefont {Opatrn{\'y}}\ \emph {et~al.}(2000)\citenamefont
  {Opatrn{\'y}}, \citenamefont {Kurizki},\ and\ \citenamefont
  {Welsch}}]{Opatrny2000}%
  \BibitemOpen
  \bibfield  {author} {\bibinfo {author} {\bibfnamefont {T.}~\bibnamefont
  {Opatrn{\'y}}}, \bibinfo {author} {\bibfnamefont {G.}~\bibnamefont
  {Kurizki}}, \ and\ \bibinfo {author} {\bibfnamefont {D.~G.}\ \bibnamefont
  {Welsch}},\ }\href {\doibase 10.1103/PhysRevA.61.032302} {\bibfield
  {journal} {\bibinfo  {journal} {Phys. Rev. A}\ }\textbf {\bibinfo {volume}
  {61}},\ \bibinfo {pages} {032302} (\bibinfo {year} {2000})}\BibitemShut
  {NoStop}%
\bibitem [{\citenamefont {Cochrane}\ \emph {et~al.}(2002)\citenamefont
  {Cochrane}, \citenamefont {Ralph},\ and\ \citenamefont
  {Milburn}}]{Cochrane2002}%
  \BibitemOpen
  \bibfield  {author} {\bibinfo {author} {\bibfnamefont {P.~T.}\ \bibnamefont
  {Cochrane}}, \bibinfo {author} {\bibfnamefont {T.~C.}\ \bibnamefont {Ralph}},
  \ and\ \bibinfo {author} {\bibfnamefont {G.~J.}\ \bibnamefont {Milburn}},\
  }\href {\doibase 10.1103/PhysRevA.65.062306} {\bibfield  {journal} {\bibinfo
  {journal} {Phys. Rev. A}\ }\textbf {\bibinfo {volume} {65}},\ \bibinfo
  {pages} {062306} (\bibinfo {year} {2002})}\BibitemShut {NoStop}%
\bibitem [{\citenamefont {Olivares}\ \emph {et~al.}(2003)\citenamefont
  {Olivares}, \citenamefont {Paris},\ and\ \citenamefont
  {Bonifacio}}]{Olivares2003}%
  \BibitemOpen
  \bibfield  {author} {\bibinfo {author} {\bibfnamefont {S.}~\bibnamefont
  {Olivares}}, \bibinfo {author} {\bibfnamefont {M.~G.~A.}\ \bibnamefont
  {Paris}}, \ and\ \bibinfo {author} {\bibfnamefont {R.}~\bibnamefont
  {Bonifacio}},\ }\href {\doibase 10.1103/PhysRevA.67.032314} {\bibfield
  {journal} {\bibinfo  {journal} {Phys. Rev. A}\ }\textbf {\bibinfo {volume}
  {67}},\ \bibinfo {pages} {032314} (\bibinfo {year} {2003})}\BibitemShut
  {NoStop}%
\bibitem [{\citenamefont {Allevi}\ \emph {et~al.}(2010)\citenamefont {Allevi},
  \citenamefont {Andreoni}, \citenamefont {Bondani}, \citenamefont {Genoni},\
  and\ \citenamefont {Olivares}}]{Allevi2010}%
  \BibitemOpen
  \bibfield  {author} {\bibinfo {author} {\bibfnamefont {A.}~\bibnamefont
  {Allevi}}, \bibinfo {author} {\bibfnamefont {A.}~\bibnamefont {Andreoni}},
  \bibinfo {author} {\bibfnamefont {M.}~\bibnamefont {Bondani}}, \bibinfo
  {author} {\bibfnamefont {M.~G.}\ \bibnamefont {Genoni}}, \ and\ \bibinfo
  {author} {\bibfnamefont {S.}~\bibnamefont {Olivares}},\ }\href {\doibase
  10.1103/PhysRevA.82.013816} {\bibfield  {journal} {\bibinfo  {journal} {Phys.
  Rev. A}\ }\textbf {\bibinfo {volume} {82}},\ \bibinfo {pages} {013816}
  (\bibinfo {year} {2010})}\BibitemShut {NoStop}%
\bibitem [{\citenamefont {Chesi}\ \emph {et~al.}(2021)\citenamefont {Chesi},
  \citenamefont {Allevi},\ and\ \citenamefont {Bondani}}]{Chesi2021}%
  \BibitemOpen
  \bibfield  {author} {\bibinfo {author} {\bibfnamefont {G.}~\bibnamefont
  {Chesi}}, \bibinfo {author} {\bibfnamefont {A.}~\bibnamefont {Allevi}}, \
  and\ \bibinfo {author} {\bibfnamefont {M.}~\bibnamefont {Bondani}},\ }\href
  {\doibase 10.3390/app11104579} {\bibfield  {journal} {\bibinfo  {journal}
  {Appl. Sci.}\ }\textbf {\bibinfo {volume} {11}} (\bibinfo {year} {2021}),\
  10.3390/app11104579}\BibitemShut {NoStop}%
\bibitem [{\citenamefont {Gao}\ \emph {et~al.}(2018)\citenamefont {Gao},
  \citenamefont {Lester}, \citenamefont {Zhang}, \citenamefont {Wang},
  \citenamefont {Rosenblum}, \citenamefont {Frunzio}, \citenamefont {Jiang},
  \citenamefont {Girvin},\ and\ \citenamefont {Schoelkopf}}]{Gao2018}%
  \BibitemOpen
  \bibfield  {author} {\bibinfo {author} {\bibfnamefont {Y.~Y.}\ \bibnamefont
  {Gao}}, \bibinfo {author} {\bibfnamefont {B.~J.}\ \bibnamefont {Lester}},
  \bibinfo {author} {\bibfnamefont {Y.}~\bibnamefont {Zhang}}, \bibinfo
  {author} {\bibfnamefont {C.}~\bibnamefont {Wang}}, \bibinfo {author}
  {\bibfnamefont {S.}~\bibnamefont {Rosenblum}}, \bibinfo {author}
  {\bibfnamefont {L.}~\bibnamefont {Frunzio}}, \bibinfo {author} {\bibfnamefont
  {L.}~\bibnamefont {Jiang}}, \bibinfo {author} {\bibfnamefont {S.~M.}\
  \bibnamefont {Girvin}}, \ and\ \bibinfo {author} {\bibfnamefont {R.~J.}\
  \bibnamefont {Schoelkopf}},\ }\href {\doibase 10.1103/PhysRevX.8.021073}
  {\bibfield  {journal} {\bibinfo  {journal} {Phys. Rev. X}\ }\textbf {\bibinfo
  {volume} {8}},\ \bibinfo {pages} {021073} (\bibinfo {year}
  {2018})}\BibitemShut {NoStop}%
\end{thebibliography}%
\end{document}